\documentclass[]{spie}  
\usepackage[]{graphicx}
\usepackage{float}
\usepackage{amssymb,amsmath}
\title{Design of a Full-Stokes Polarimeter for VLT/X-shooter} 

\author{Frans Snik\supit{a}, Gerard van Harten\supit{a}, Ramon Navarro\supit{b}, Paul Groot\supit{c}, Lex Kaper\supit{d}, 
Alfred de Wijn\supit{e}\skiplinehalf
\small{\supit{a}Sterrewacht Leiden, Universiteit Leiden, Niels Bohrweg 2, 2333 CA, Leiden, the Netherlands; \\
\supit{b}NOVA Optical Infrared Instrumentation group at ASTRON, PO Box 2, 7990 AA Dwingeloo, The Netherlands;\\
\supit{c}Radboud Universiteit Nijmegen, Postbus 9010, 6500 GL, Nijmegen, The Netherlands; \\
\supit{d}Astronomical Institute ``Anton Pannekoek'', University of Amsterdam, P.O. Box 94249, 1090 GE, Amsterdam, The Netherlands;\\
\supit{e}High Altitude Observatory, NCAR, PO Box 3000, Boulder, CO, 80307-3000, USA.}
}

\authorinfo{email: {\tt snik@strw.leidenuniv.nl}}

\begin{document} 
  \maketitle 

\begin{abstract}
X-shooter is one of the most popular instruments at the VLT, offering instantaneous spectroscopy from 300 to 2500 nm. 
We present the design of a single polarimetric unit at the polarization-free Cassegrain focus that serves all three spectrograph arms of X-shooter. 
It consists of a calcite Savart plate as a polarizing beam-splitter and a rotatable crystal retarder stack as a ``polychromatic modulator''. 
Since even ``superachromatic'' wave plates have a wavelength range that is too limited for X-shooter, this novel modulator is designed to offer close-to-optimal polarimetric efficiencies for all Stokes parameters at all wavelengths. 
We analyze the modulator design in terms of its polarimetric performance, its temperature sensitivity, and its polarized fringes. 
Furthermore, we present the optical design of the polarimetric unit. 
The X-shooter polarimeter will furnish a myriad of science cases: from measuring stellar magnetic fields (e.g., Ap stars, white dwarfs, massive stars) to determining asymmetric structures around young stars and in supernova explosions.
\end{abstract}



\section{SCIENCE CASES FOR AN X-SHOOTER POLARIMETER}
Polarization is an often neglected carrier of important physical information contained in the radiation field of celestial objects.
Polarization studies reveal detailed and unique knowledge on e.g.~the geometry (asymmetry) of the radiating source, the strength and orientation of magnetic fields, and the radiation mechanism (synchrotron, cyclotron emission). 
No astrophysical object is perfectly spherically symmetric; light from these objects is therefore never completely unpolarized, and measuring that polarization, however small, in spectral lines and/or the continuum provides crucial information on the physical environment within which the light we observe was produced.

As X-shooter measures the continuum and spectral lines from the UV (300 nm) up to the near-IR (2500 nm), the science cases motivating a spectropolarimetric capability for X-shooter are therefore widespread.
And as some information reaches us in the form of linear polarization, and some other information in the form of circular polarization, a full-Stokes polarimeter in conjunction with X-shooter will constitute a truly powerful instrument. 
Below we highlight a selection of these science cases\cite{SnikKellerreview,Xshooterpolcases} to be carried out with the X-shooter spectropolarimeter.

\subsection{Stellar Magnetic Fields}

Spectropolarimetry has completely revolutionized our dynamical view of stars.
Recent results\cite{stellarmagnetism, FORSlegacy,Mdwarfs, MiMeS} show that stars of all types can harbor (surface) magnetic fields, from fully convective dwarfs to very massive stars having predominantly radiative envelopes. 
The dynamo mechanisms generating the magnetic fields are likely different from that of solar-type stars, and all are still poorly understood. 
However, stellar magnetic fields have a dramatic impact on the circumstellar environment, and, perhaps even more importantly, on the evolution of the star itself.

Stellar magnetic fields are directly measured through the Zeeman effect, acting on most of the stellar spectral lines. 
A magnetic field along the line-of-sight produces an antisymmetric line pattern in circular polarization (Stokes $V$), whereas a transverse magnetic field leaves a symmetric imprint in linear line polarization (Stokes $Q$, $U$). 
However, the degree of polarization for spectral lines of averaged starlight is typically small in Stokes $V$ ($<10^{-3}$), and usually even an order of magnitude smaller in $Q$ and $U$.
For most stars, Zeeman effect signals cannot be measured in individual spectral lines, as they drown in photon noise. 
But all the hundreds of spectral lines within a certain wavelength range can be added in an appropriate way\cite{LSD1,LSD2}, such that a magnetic field measurement can still be carried out. 
Magnetic fields as small as 0.1 Gauss can be detected with current instrumentation\cite{HARPSpol2}. 
Moreover, by measuring the Zeeman signals as a function of the stellar rotation, maps of the stellar magnetic field can be inferred\cite{ZDI,MDI} for e.g.~Ap and T~Tau stars, even despite the fact that the star remains a point source. 
Magnetic fields have thus been detected and mapped for stars throughout the HR diagram\cite{Mdwarfs, MiMeS}, including protostars and white dwarfs\cite{FORSlegacy}. 
Because  Zeeman splitting scales with $\lambda^2$, whereas line broadening scales with $\lambda$, there is a clear opportunity for carrying out spectropolarimetric measurements in the IR, particularly for cool stars.

The edge of a stellar disk is generally polarized parallel to the limb in continuum and spectral lines, with largest fractional polarization in the blue and UV. 
Because of symmetry, this polarization disappears in integrated starlight. 
However, a planetary transit can break this symmetry\cite{limbpoltransit}, providing a means to both detect exoplanets and constrain their orbits, as well as diagnose the stellar atmosphere. 
The symmetry can also be broken by a weak dipolar field, which modifies the line polarization through the Hanle effect\cite{stellarHanle}. 
Line polarization also occurs in optically pumped wind lines, and this polarization is modified by magnetic realignment.\cite{UVpol}

\subsection{Cyclotron \& Synchrotron Radiation produced by Relativistic Blast Waves}

Both cyclotron and synchrotron emission are strongly polarized in the continuum, also in the UV--NIR range. 
The (phase-averaged) radiation from high-energy sources as pulsars\cite{pulsarpolarization} and AGN jets\cite{AGNjetpol} can therefore be polarimetrically analyzed to infer their magnetic field structure. 
In the case of a long gamma-ray burst (GRB) afterglow, i.e.~a relativistic blast wave plowing through the circumstellar medium of the collapsed massive star, the detected radiation is consistent with synchrotron emission characterized by a series of smoothly connected power laws with characteristic break frequencies and fluxes. 
Although the macroscopic properties of these relativistic shocks are largely understood, the acceleration of the relativistic particles and the physical origin and structure of the magnetic field is still a mystery. 
The magnetic field present in the blast wave can be probed by circular polarization, and the variability in geometry by linear polarization. 
This has been done for a few very bright GRB afterglows, but as these afterglows are only not very strongly polarized (a few percent\cite{Greiner, GRB1,Wiersema2012, GRB2}), a very high signal to noise ratio is necessary, requiring an 8-m class telescope. 
For short GRBs, polarimetry may be the most powerful tool to identify the progenitor: in the binary merger model, (one of) the neutron stars is a non-recycled pulsar, having a powerful magnetic field. 
This could perhaps be captured in the blast wave, producing polarization values higher than those of long GRBs.

\subsection{Asymmetric Supernovae}

Recent observations indicate that supernovae are not round. 
For nearly a century it has been assumed that supernovae are more or less spherical, since there was no strong evidence to the contrary. 
Only in the last few years has the realization dawned that this assumption may be fundamentally wrong -- that supernovae are not spherically
expanding, but, like proposed in the fireball model explaining GRBs, produce lopsided, squirted jets -- a finding that has deep implications.

Spectropolarimetric observations of supernovae\cite{SN} have shown that their light is polarized at the 1--2~\% level, with prominent differences between different spectral lines, and thus providing important information on the distribution of different chemical elements produced in the remnant. 
There are, in principle, lots of reasons why the light from a supernova could be polarized. 
The supernova could be out of round, it could be round but have off-center sources of light, or other matter in the vicinity could be asymmetrically distributed. 
It turns out that supernovae of type Ia, which are supposed to be produced by white dwarfs pushed over the Chandrasekhar limit through accretion of matter from a binary companion, show little or no polarization signal. 
Supernovae that are thought to arise by core collapse in massive stars (Type II, Ib, and Ic) are, however, all polarized. 
This conclusion is still based on small-number statistics, as spectropolarimetry requires large apertures, even for sources as bright as supernovae.

\subsection{Scattering Disks \& Atmospheres}
Like our blue sky, any volume of gas or dust can linearly polarize starlight upon scattering.
Measuring stellar continuum polarization therefore yields information on the presence of asymmetric circumstellar structures as protoplanetary disks. 
Continuum and emission line spectropolarimetry yields direct information on the disk structure and rotation rate.\cite{Vink}
Absorption lines of Herbig Ae/Be stars have been found to be linearly polarized\cite{Harrington}, which is attributed to selective absorption of the optically pumped gas in the disk\cite{Kuhnpumping}.

The combined light of the star and scattered light off an exoplanetary atmosphere may measurable\cite{Berdyugina}, particularly in the blue/UV where scattering is most prominent.
Scattering also takes place in the hazy atmospheres of brown dwarfs\cite{browndwarfs}, and non-zero polarization hints at stellar oblateness or even patchy clouds.

The light scattered of the accretion disk in interacting binaries is generally polarized, and its polarization variation with time allows for the determination of the disk inclination and physical properties.\cite{interacting binaries}

\subsection{AGNs}
Spectropolarimetric observations have led to the unification of type 1(pole-on)  and type 2 (edge-on) active galactic nuclei\cite{AGNunification}, as the polarized spectral flux allows for a direct around-the-corner peer into the nucleus itself, even though it is obscured from view.
The spectral variation of the continuum polarization yields information on the clumpiness of the scattering medium.
Spectral line polarization of AGNi yields vital clues about the circumnuclear structure and kinematics.\cite{AGNspectropol}

\subsection{Opportunities for an X-shooter Polarimeter}
X-shooter\cite{Xshooter} is a unique instrument as it combines a tremendous wavelength range (300--2500 nm) with unprecedented efficiency and medium-range spectral resolution, and is hence a very highly demanded instrument.
These properties would become even more important in case X-shooter is equipped with a spectropolarimetric mode.
In comparison with current spectropolarimeters at 8-m class telescopes (VLT/FORS\cite{FORS1, PatatRomaniello, FORSpipeline, FORSlegacy}, Subaru/FOCAS\cite{SubaruFOCAS} and Keck/LRISp\cite{KeckLRISp}, which all operate in the visible range), X-shooter-pol will have a much larger wavelength coverage, and improved spectral resolution.
The SALT/RSS\cite{SALTRSSpol} spectropolarimeter is being commissioned, and goes down to 320 nm, but is hampered by instrumental polarization issues. 
The discovery space for spectropolarimetry below 400 nm is untouched and likely rich.
Also near-IR spectropolarimetry is severely underexploited, particularly at 8-m class telescopes.
Often, different spectral lines probe different regions in the stellar atmosphere/circumstellar environment, and therefore a wide wavelength range is required to yield a complete picture.

High-resolution spectropolarimeters like CFHT/ESPaDOnS\cite{ESPaDOnS} and ESO 3.6-m/HARPSpol\cite{HARPSpol1, HARPSpol2, HARPSpol3} (both operate in the visible range) have brought us revolutionary results on stellar magnetism.
Compared to X-shooter these instruments have larger spectral resolutions by a factor of $\sim$10.
However, most spectral lines, particularly of hot stars, will still be resolved by X-shooter.
Moreover, owing to the telescope's larger collection area (8.2-m as compared to 3.6-m) and the spectrographs' larger efficiencies and smaller spectral resolution, many more magnetic stars will be accessible to X-shooter-pol.
Because both ESPaDOnS and HARPS are fiber-fed, they cannot measure continuum polarization (because of the variable transmissions of the fibers during tracking), whereas the slit spectrographs of X-shooter in principle allow for the measurement of continuum polarization.

Several near-IR high-resolution spectropolarimeters are currently being designed (CFHT/SPIRou\cite{SPIRou}, VLT/ CRIRES upgrade\cite{CRIRES-pol}, GTC/MIRADAS \cite{MIRADAS}), and will specifically address magnetism in young and cool stars.
By far the most powerful property of X-shooter-pol will be the instantaneous full-Stokes observation in the UV, visible and near-IR ranges, for a wide range of targets.

\section{REQUIREMENTS}
To address the science goals introduced in the previous section, we distill the following requirements for a polarimetric unit for X-shooter:
\begin{itemize}
\item All three polarized \textit{Stokes parameters} ($Q,U$ for linear polarization; $V$ for circular polarization) shall be measurable.
\item The full \textit{spectral range} of X-shooter (300--2500 nm) shall in principle be covered by the polarimeter. If necessary, sacrifices can be made at the edges, as not many photons are transmitted by the Earth's atmosphere from 300--350 nm, and stray-light issues limit X-shooter's performance in the K-band.
\item Polarimetric observations shall be enabled for both \textit{spectral lines} and the \textit{continuum}.
\item The \textit{polarimetric sensitivity}\cite{SnikKellerreview} (for line polarization) shall not be limited by systematic effects down to absolute fractional polarizations $< 10^{-4}$.
\item The measured Stokes parameters are related to the real Stokes parameters through the $4\times4$ response matrix $\sf{X}(\lambda)$:
\begin{equation}
{\sf X}(\lambda)= \left( \begin{array}{cccc} I\rightarrow I & Q\rightarrow I & U\rightarrow I & V\rightarrow I \\ I\rightarrow Q & Q\rightarrow Q & U\rightarrow Q & V\rightarrow Q \\ I\rightarrow U & Q\rightarrow U & U\rightarrow U & V\rightarrow U \\ I\rightarrow V & Q\rightarrow V & U\rightarrow V & V\rightarrow V  \end{array} \right)(\lambda)\, .
\end{equation}
This matrix $\sf{X}(\lambda)$ shall include all effects of data reduction and calibration, but excludes the effects of random noise on the data (such as photon and rad-ut noise, and the second-order effects of seeing and residual instrument drifts).
The \textit{polarimetric accuracy}\cite{SnikKellerreview} for the X-shooter polarimeter is then required to be:
\begin{equation}
\Delta {\sf X}(\lambda) =  {\sf X}(\lambda) - \mathbb{I} < \left( \begin{array}{cccc} - & 0.1 & 0.1 & 0.1 \\ 0.0001 & 0.01 & 0.01 & 0.01\\0.0001 &0.01 & 0.01 & 0.01 \\ 0.001 & 0.01 & 0.01 & 0.01  \end{array} \right)\,.
\end{equation}
The first column of $\Delta {\sf X}$ pertains to the \textit{instrumental polarization} which limits the measurement of a small continuum polarization in $Q$, $U$ or $V$ as it varies the zero point of the polarization measurement.
The $3\times3$ block on the lower block describes \textit{polarization rotation} and \textit{cross-talk}, which affect the scaling of the polarization measurement, and could mask a small signal by a cross-talk signal from another Stokes parameter.
\item The \textit{polarization efficiencies}\cite{poleff} with which $Q$, $U$ and $V$ are measured shall be close to their optimal values ($1/\sqrt{3}$) for all wavelengths within the specified range, and certainly $>0.5$.
\item The \textit{transmission} of the polarization optics train shall be $> 70\%$ for all wavelengths within the specified range.
\end{itemize}

We adopt the following assumptions and boundary conditions for the design of the polarimetric unit for X-shooter:
\begin{itemize}
\item The unit will be located at the Cassegrain focus, which is essentially free from instrumental polarization induced by the telescope. The optics are therefore fed with an F/13.4 beam, and the mechanical design has to take the variable gravity vector into account.
\item The unit is to be mounted onto the Acquisition \& Guiding sliding mechanism, such that it is fully retractable. A volume of $\sim 80 \times 80 \times 80$ mm$^3$ would reasonably be allotted.
\item A return beam of the peripheral field is to be supplied to the guiding camera.
\item Calibration needs to be guaranteed within the accuracy requirements for a temperature range of $\pm5$ $^\circ$C. The full temperature range is 0--30 $^\circ$C.
\item As the ADCs are implemented downstream the Cassegrain focus, the polarimetric unit shall fully transmit an atmospherically dispersed stellar image.
\item The slits of all three spectrographs are to be fed with polarimetrically analyzed light. Slit dimensions of 1''$\times$12'' are assumed. The three slits can be aligned individually.
\item A provision for nodding inside the IR spectrograph needs to be implemented.
\item The split linear polarization directions should be at $\pm 45^\circ$ with respect to the dichroic splitting filters and the gratings, to ensure that no differential spectral transmission effects occur between the two beams.
\end{itemize}

\section{POLYCHROMATIC MODULATOR}
The most major challenge for the X-shooter polarimeter is to design a polarization modulator that covers the enormous wavelength range of 300--2500 nm (more than three octaves!).
The common solution which is applied in all current spectropolarimeters is to implement achromatic or ``superachromatic'' wave plates.
These modulators are either half-wave (for linear polarimetry) or quarter-wave (for circular polarimetry), and their bandwidths are increased by combining crystals with different dispersions of their birefringence\cite{Beckers} and/or Pancharatnam\cite{Pancharatnam} combinations of several plates.
Even the superachromatic plates cover only slightly more than one octave, and two of them would be required to cover the wavelength range of X-shooter.
Fresnel rhombs could be tailored to cover such a large wavelength range\cite{ESPaDOnS}, but the limited volume in the Cassegrain focus of the X-shooter instrument prohibits the use of those.
Moreover, the limited volume even prohibits the implementation of two wave plates (half-wave and quarter-wave) to furnish full-Stokes measurements.

All these considerations led us to adopt a radically different approach to polarization modulation.
Instead of achromatizing the modulation itself, we choose to achromatize the polarization efficiencies\cite{poleff} for $Q$, $U$ and $V$ at every wavelength within the range.
These efficiencies indicate to what degree the measurable Stokes parameters can be demodulated with respect to the noise, given a certain modulation scheme.
For optimal efficiencies, the error propagation of the noise to the fractional polarizations $Q/I$, $U/I$, $V/I$ will be most benign.
Let the polarization modulation process of $n$ states be described by the $n \times 4$ modulation matrix ${\sf O}(\lambda)$\cite{SnikKellerreview}:
\begin{equation}
\mathbf{I}'(\lambda)= {\sf O}(\lambda) \cdot \mathbf{S}_{\textrm{\scriptsize{in}}}(\lambda)\,.
\label{eqmodulation}
\end{equation}
The column vector $\mathbf{I}'(\lambda)$ contains the $n$ spectral measurements according to the $n$ modulation states $i$.
The $n$ rows of the modulation matrix {\sf O} are in fact the first rows of the total instrumental Mueller matrix for the different modulation states $i$.
These rows describe the conversion of the incident Stokes vector $\mathbf{S}_{\textrm{\scriptsize{in}}}$ to the intensity spectra $I'_i(\lambda)$ that is measured by the detector.
In our case, the modulation is effectuated by the six-step rotation of a stack of crystals that is to be optimized.
As the polarimetric unit is to be located in a polarization-free Cassegrain focus before any other optics than M1 and M2, the Mueller matrices of the modulator and the polarizing beam-splitter (analyzer) are the only ones to be considered in ${\sf O}(\lambda)$.
As ${\sf O}(\lambda)$ is a $6\times4$ matrix, the Stokes vector is (in principle) overdetermined by this measurement.
The most optimal demodulation matrix\cite{poleff} is the Moore-Penrose pseudo-inverse of  ${\sf O}(\lambda)$:
\begin{eqnarray}
\mathbf{S}_{\textrm{\scriptsize{out}}}(\lambda) & = & {\sf D}(\lambda) \cdot \mathbf{I}'(\lambda)\,;\\
{\sf D}(\lambda)  & = & ({\sf O}^T(\lambda)  {\sf O}(\lambda) )^{-1} {\sf O}^T(\lambda)  \,.
\end{eqnarray}
This optimal demodulation maximizes the combined polarization efficiencies, defined as:
\begin{equation}
\epsilon_k = \left(n\sum_{l=1}^n {\sf D}_{kl}^2 \right)^{-\frac{1}{2}}\,.
\label{OptPolEff}
\end{equation}
These efficiencies obey:
\begin{equation}
\epsilon_1 \le 1 \, , \qquad \sum_{k=2}^4 \epsilon_k^2 \le 1\,.
\end{equation}
Therefore, for an optimally balanced full-Stokes modulator, the polarization efficiencies for $Q$, $U$, and $V$ are $1/\sqrt{3}$.
This number is to be compared with efficiencies of $1/\sqrt{2}$ for $Q$ and $U$ with a step-wise rotating half-wave plate polarimeter, and with an efficiency of 1 for $V$ with a rotating quarter-wave plate polarimeter.

As the X-shooter polarimeter is required to measure all the Stokes parameters, we optimize a ``polychromatic modulator''\cite{polychromatic, polychromatic2, polychromatic3} that offers close-to-optimal ($\approx 1/\sqrt{3} \approx 0.58\%$) efficiencies for $Q$, $U$ and $V$ at all wavelengths within the range of 300--2500 nm.
The design trade-off is described in the following subsections.

It is well-known that a dual-beam polarimeter\cite{ZDI,Bagnulodualbeam} is crucial to deal with the time-variable signals due to seeing.
For rotating half-wave and quarter-wave plate polarimeters, the dual-beam demodulation can be performed through a double difference and a double ration.
In both cases, the systematic effects due to seeing and due to differential effects between the two beams cancel to first order.
Fortunately, the double difference demodulation is also fully applicable to polychromatic modulation\cite{generaldualbeam, SnikKellerreview}:
\begin{equation}
\frac{\mathbf{S}_{\textrm{\scriptsize{out}}}(\lambda)}{I_{\textrm{\scriptsize{out}}}(\lambda)} = \frac{1}{2} \left[ \frac{\mathbf{S}_{\textrm{\scriptsize{out}},L}(\lambda)}{I_{\textrm{\scriptsize{out}},L}(\lambda)} - \frac{\mathbf{S}_{\textrm{\scriptsize{out}},R} (\lambda)}{I_{\textrm{\scriptsize{out}},R}(\lambda)} \right] \,.
\end{equation}

\subsection{Design Trade-off}
We have explored several solutions for the X-shooter polychromatic modulator\cite{Xshooter-modulator}.
Firstly, we identified quartz (SiO$_2$), MgF$_2$ and sapphire (Al$_2$O$_3$) as the only  viable birefringent materials that are transparant over the entire wavelength range.
Next, several ``unit stacks'' were designed from which the compound modulator will be constructed.
These unit stacks can have several properties:
 \begin{itemize}
 \item Quasi zero-order (little retardance change with temperature);
 \item Achromatic combination of two different crystals\cite{Beckers} (less unit stacks needed for polychromatic modulation);
 \item Athermal combination of two different crystals\cite{HaleDay} (minimum retardance change with temperature);
 \item Wide-field\cite{HaleDay} (minimum retardance change at non-normal angle of incidence);
 \item A compromise between the above properties.
 \end{itemize}

The numerical optimization of the modulation scheme is performed using a simulated annealing algorithm\cite{polychromatic}.
This allows for crossing a local optimum (with a probability depending on the goodness of the optimum, and less crossings are accepted as the simulations are progressing), such that in the end, ideally, the system is frozen in the global optimum.
Because for virtually all astronomical sources $I \gg Q,U,V$, it is not very relevant to optimize the efficiency for $I$ for its demodulation to be independent of $Q$, $U$ and $V$.
Hence, during optimization, we keep track of two figures of merit:
\begin{eqnarray}
\nonumber \Delta\epsilon_a \equiv \left< \sqrt{ \left(\frac{1}{\sqrt 3}-\epsilon_Q\left(\lambda\right)\right)^2 + \left(\frac{1}{\sqrt 3}-\epsilon_U\left(\lambda\right)\right)^2 + \left(\frac{1}{\sqrt 3}-\epsilon_V\left(\lambda\right)\right)^2} \right>_\lambda, \\
\label{eq:merit1}
\end{eqnarray}
where the $\left<\cdots \right>_\lambda$ denotes spectral averaging, and:
\begin{eqnarray}
\Delta\epsilon_b \equiv \max\left( 1 - \sqrt{3}\min\left(\epsilon_Q\left(\lambda\right), \epsilon_U\left(\lambda\right), \epsilon_V\left(\lambda\right) \right) \right).
\label{eq:merit2}
\end{eqnarray}
The eventual merit function to be minimized is a linear combination of these two figures of merit, which can be tuned to the desired type of solution.
Minimization of the condition number of the modulation matrix\cite{polychromatic2,polychromatic3} gives similar results, but it lacks the handle on the efficiency for individual Stokes parameters.
As a waveplate's retardance scales with $1/\lambda$, efficiency variations are typically periodic in the inverse wavelength domain.
In order to avoid undersampling and thus low efficiencies at small wavelengths, the spectral sampling of the figures of merit is equidistant in the frequency domain.

The design goal for our polychromatic modulator is a maximum spectrally averaged efficiency loss of 5\%, i.e. $\Delta \epsilon_a \leq 0.05$, with a maximum monochromatic efficiency loss of 20\%, i.e. $\Delta \epsilon_b \leq 0.20$ or equivalently $\epsilon_{Q,U,V}\left( \lambda \right) \geq 0.80/\sqrt{3} \approx 0.46$.
This goal should be met whilst taking into account:
\begin{itemize}
\item Operational temperature range of 0--30~$^\circ$C;
\item F/13 beam angles of incidence;
\item Maximum thickness of the modulator of $\sim$ 15 mm;
\item Polarized spectral fringes to within the accuracy requirements;
\item Maximized transmission;
\item Manufacturing tolerances.
\end{itemize}

After have explored the design space, we down-selected the unit stacks to:
\begin{enumerate}
\item Quasi zero-order quartz\cite{Dodge1984}; thin unit stack;
\item Quasi zero-order quartz; thick unit stack;
\item Achromatized MgF$_2$\cite{Ghosh1999} -- quartz combination;
\end{enumerate}
Based on each unit stack, a polychromatic modulator was optimized.
The results of the trade-off between these designs are:
  \begin{table}[H]
\centering
$
        \begin{array}{p{0.40\linewidth}|p{0.10\linewidth}|p{0.10\linewidth}|p{0.10\linewidth}}
           \hline
           \noalign{\smallskip}
           Design & 1 & \textbf{2} & 3 \\
           \noalign{\smallskip}
           \hline
           \hline
           \noalign{\smallskip}
           Efficiencies for all wavelengths 		& + & \textbf{+} & + \\
           Chromaticity of the modulation matrix		& + & \textbf{+} & ++ \\
           Polarized fringes			& $-$ & \textbf{+} & $- -$ \\
           Temperature effects		& $-$ & $\mathbf{-}$ & $- -$ \\
           Off-axis variations			& + & \textbf{+} & + \\
           Manufacturability	& $- -$ & $\mathbf{-}$ & + \\
           \noalign{\smallskip}
           \hline
        \end{array}
$
  \end{table}

\begin{figure}[p]
\centering
\includegraphics[width=0.2\textwidth]{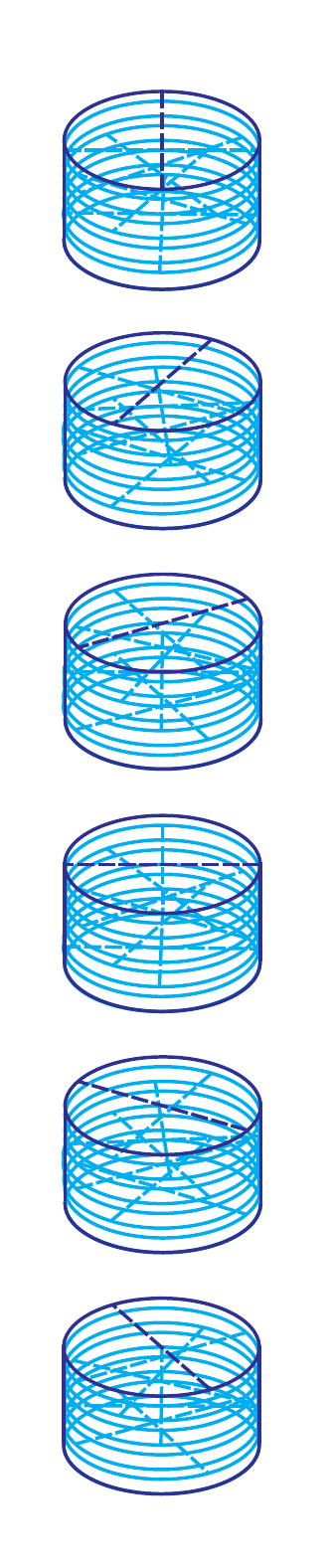}
\includegraphics[width=0.79\textwidth]{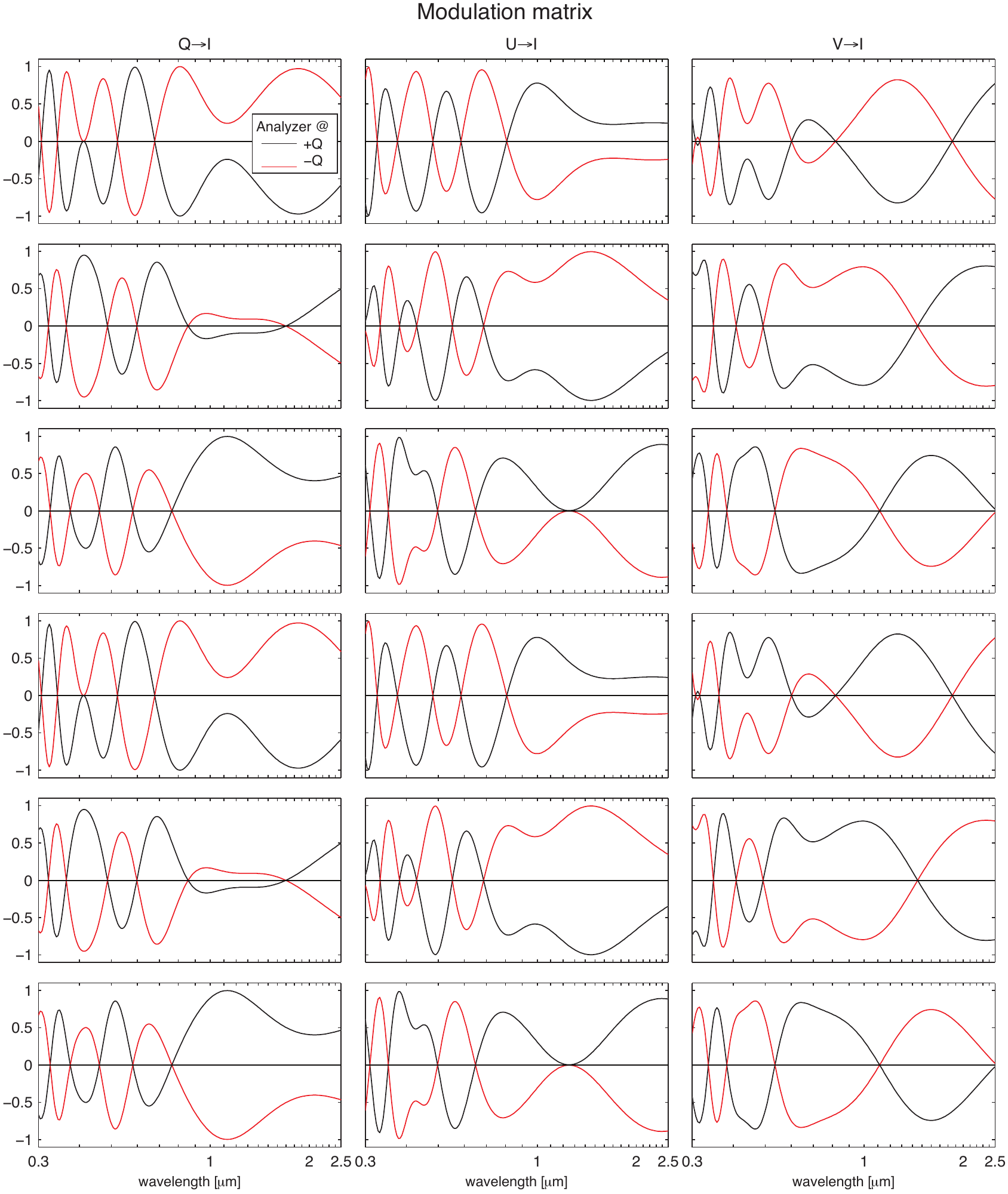}
\caption{The modulation matrix as a function of wavelength for the baseline polychromatic modulator consisting of four quasi zero-order quartz plates with different thicknesses and at different angles. The first column of matrix elements pertaining to just Stokes $I$ has been omitted.}
\label{modulationmatrix}
\end{figure}

\begin{figure}[h]
\centering
\includegraphics[width=0.6\textwidth]{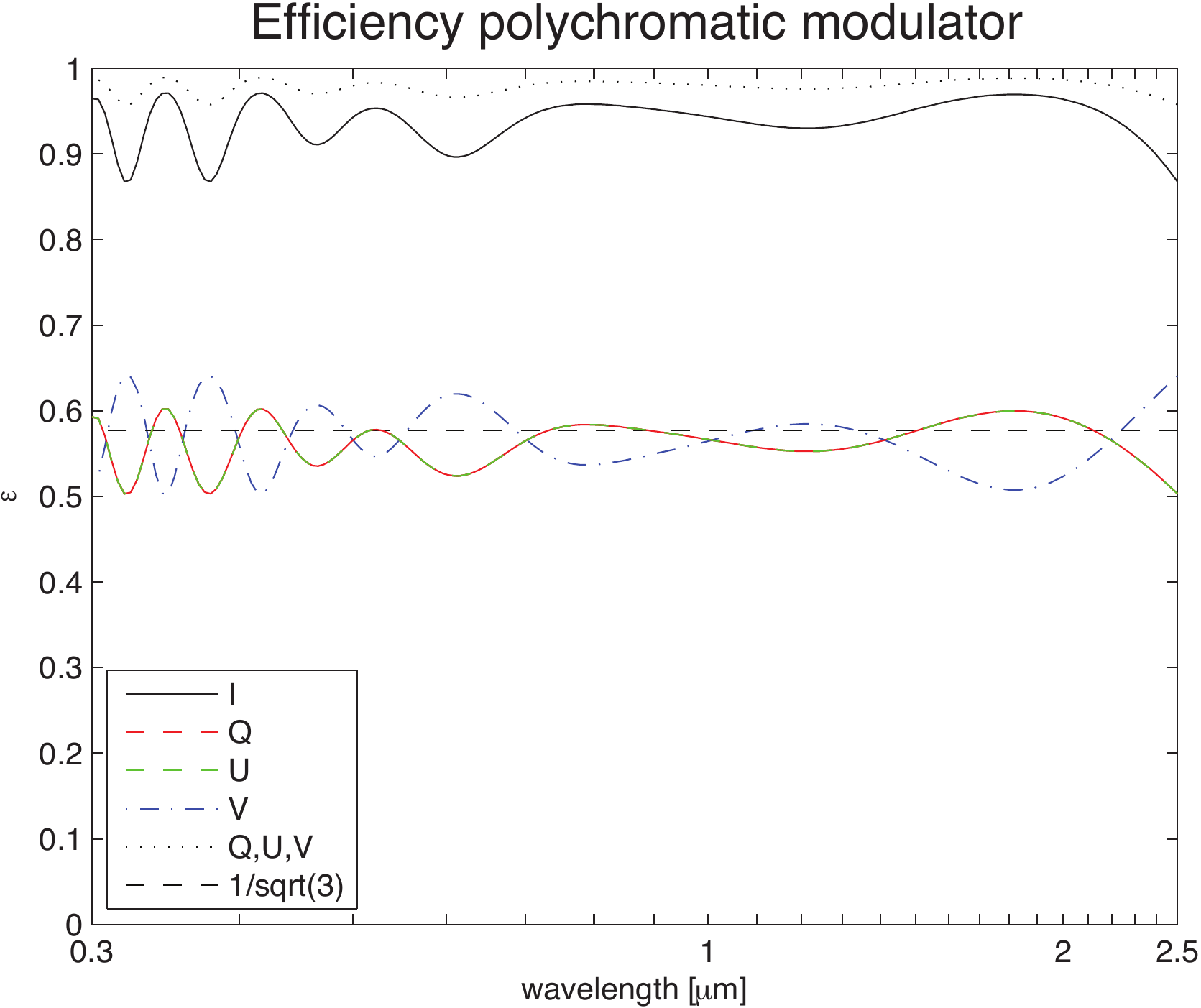}
\caption{The close-to-optimal polarization efficiencies of the baseline polychromatic modulator.}
\label{efficiencies}
\end{figure}

Based on this trade-off, design \#2 with four quasi zero-order quartz stacks (i.e.~a total of eight quartz plates with different thicknesses and angles) has been adopted as the current baseline.
The stack is to be rotated stepwise to angles [0, 30, 60, 90, 120, 150]$^\circ$.
Its modulation matrix for both beams out of the polarizing beam-splitter is presented in Fig.~\ref{modulationmatrix}, and the resulting polarization efficiencies as a function of wavelength are presented in Fig.~\ref{efficiencies}.
It is clear that the modulation matrix is very chromatic, but that the polarization efficiencies are not, and, moreover, are close-to-optimal.

A complete tolerance analysis for the manufacturing of this modulator is left for future work\cite{Xshooter-modulator}.
It is likely that for individually polished quartz plates, the manufacturing errors add up in an intolerable way.
Therefore, we aim to measure the retardance of each plate after polishing, and update the design before the next plate is polished.\cite{Mahler2011}

We are currently still investigating the design of a reconfigurable polychromatic modulator.
Often, the polarization of a target has $Q,U\gg V$ (in cases of scattering), or $V \gg Q,U$ (for the Zeeman effect).
In such cases, one would like to only demodulate to $Q$ and $U$, or to $V$. 
As potential cross-talk issues are then of little concern, the demodulation can be performed to just a subset of the Stokes vector, and ${\sf O}(\lambda)$ may be considered an $n\times3$ or $n\times2$ matrix.
The optimal demodulation matrix is then still given by the pseudo-inverse.
The demodulation to, e.g, Stokes $V$ then does not put $Q$ and $U$ in the null-space\cite{nullspace}, as they can be assumed small enough.
This way, a modulation scheme for $n<6$ steps may be found for both $Q, U$ separately and $V$ separately, which each have efficiencies $> 1/\sqrt{3}$.

\subsection{Temperature Effects}
\begin{figure}[p]
\centering
\includegraphics[angle=90, width=0.9\textwidth]{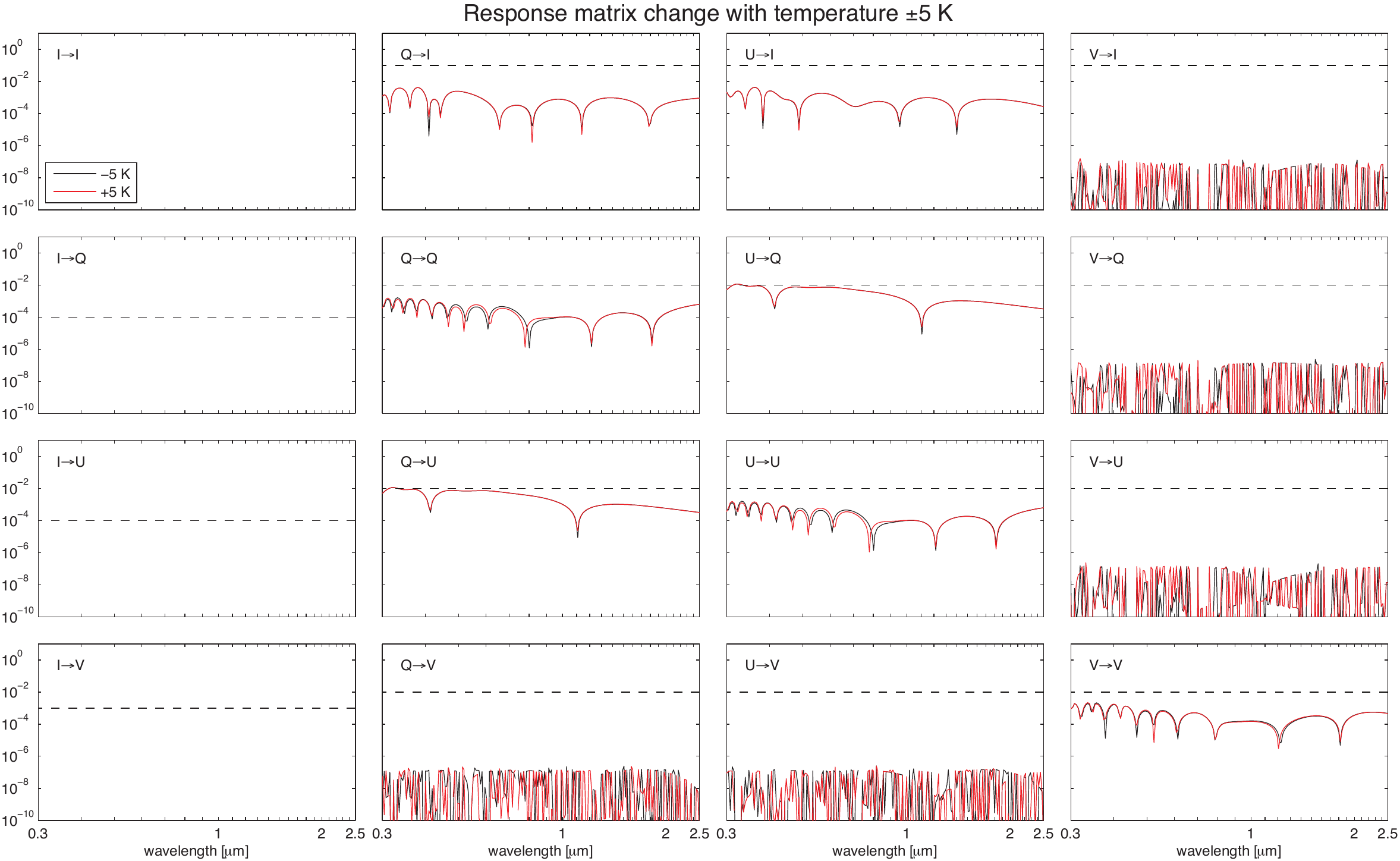}
\caption{The errors upon the modulation matrix for a 5$^\circ$ temperature change for the baseline polychromatic modulator. The dashed lines indicate the accuracy requirements.}
\label{temperrors}
\end{figure}

The retardance of a birefringent plate is very dependent on temperature, as the plate's thickness changes and also its birefringence.\cite{HaleDay}
These effects have been modeled with accurate thermo-optic data\cite{Ghosh1998} for all polychromatic modulator designs.
For the baseline design, the temperature effects are relatively small, as the units stacks are all quasi-zero order.
The errors in ${\sf X}(\lambda)$ due to a temperature change of $\pm$ 5 $^\circ$C are presented in Fig.~\ref{temperrors} (after applying the same demodulation matrix as before).
It is concluded that these errors are within the tolerances, although a recalibration may be required for larger temperature variations.

\begin{figure}[p]
\centering
\includegraphics[angle=90, width=0.9\textwidth]{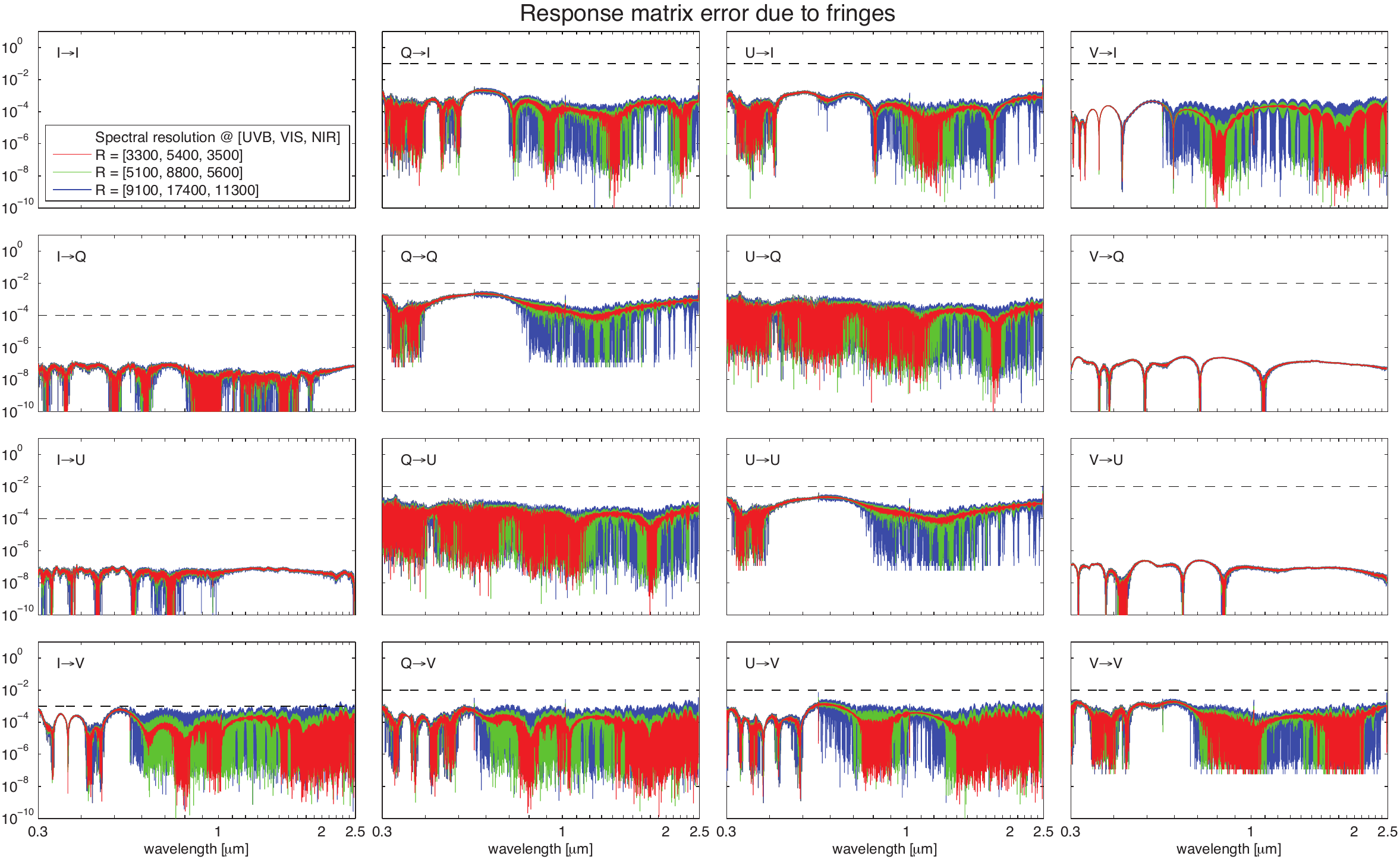}
\caption{The errors from (uncorrected) polarized fringes for the baseline polychromatic modulator. The dashed lines indicate the accuracy requirements.}
\label{fringeserrors}
\end{figure}

\subsection{Polarized Fringes}
Birefringent plates generally exhibit polarized spectral fringes\cite{fringes1,fringes2,fringes3,fringes4}.
The fringes are due to Fabry-Perot etaloning, and they are polarized, as the fringes for the polarization directions along the fast and slow axis of the birefringent plate are spectrally displaced, as they see a different refractive index.
The polarized fringes usually limit the detectability of a polarization signal in the continuum, and even in spectral lines.
For spectrographs with limited spectral resolution and for thick crystal plates, the amplitude of the polarized fringes is often decreased by spectral smearing, such that it is below the instrument's sensitivity limit (as is the case for FORS).
For high-resolution spectropolarimeters, the fringe period can be made small enough by using Fresnel rhombs\cite{ESPaDOnS} or zero-order plastic retarders\cite{HARPSpol1, HARPSpol2}, for which the fringe amplitudes are also smaller to begin with because of the small retardances involved.
Unfortunately, Fresnel rhombs would not work for the X-shooter polarimeter for reasons explained above, and the plastic retarders are not transparent over the entire wavelength range.
It is therefore clear that the effects of polarized fringes need to be carefully analyzed for the adopted polychromatic modulator, and several mitigating actions can be taken.

To minimize polarized spectral fringes, it is necessary to minimize reflections at the interfaces, so large transitions in refractive index should be avoided.
Besides cementing the plates with index-matching glue, anti-reflection coating must be applied at the outer surfaces, as there the largest transition occurs.
As we do not know the recipe of a multilayer AR coating that is applicable to the entire wavelength range, we model a coating of a few hundred nanometers of the lower refractive index material MgF$_2$ on the higher index material quartz.

The polychromatic solution requires quasi zero-order unit stacks, because of too small retardances, resulting in a doubling of the amount of plates.
This however enables us to investigate the performance for different unit stack thicknesses, such that the spectral fringes are sufficiently smeared out.

The polarized spectral fringes of the baseline modulator composed of eight $\sim$1 mm thick quartz plates including the effects of the MgF$_2$ coating were computed using Berreman calculus\cite{fringes4}.
The results were convolved with the different spectral resolutions of the three individual X-shooter spectrograph arms.
Fig.~\ref{fringeserrors} shows the effects of the fringes with respect to the response matrix ${\sf X}$ the idealized spectral modulator without internal reflections.
For all possible slits and therefore spectral resolutions, the polarized fringes are within the accuracy requirements.
The polarized fringes that cause $I \rightarrow Q$ and $I \rightarrow U$ are almost completely suppressed.
This is thanks to the optimal demodulation scheme: polarized fringes change their sign upon a 90$^\circ$ rotation of the stack.
The same holds for Stokes $V$ modulation.
The optimal demodulation for $Q$ and $U$ implies that $V$ is put in their null-space\cite{nullspace}, and therefore also the fringes are put in the null-space.
The fringes are therefore still apparent in $I \rightarrow V$.
As the demodulation is overdetermined, a specific demodulation may be found for $V$ that puts the fringes in the null-space.
The fringes amplitudes will be suppressed further by smearing due to the F/13 beam (this calculation is only for normal incidence), and upon applying a better AR coating.

\subsection{Calibration}
The performance of the polychromatic modulator will be measured already during production, and in the lab directly afterwards.
The modulation matrix elements are to be measured (at least for a major part of the wavelength range) by supplying a set-up with the step-wise rotating modulator and a polarizer with light of known polarization.
Stokes $Q$ and $U$ are obviously created with a rotating polarizer, and Stokes $V$ is created by adding a rotating quarter-wave plate.
Such a calibration unit can be used to infer the modulation matrix through a least-squares method, whilst also determining the errors of the calibration optics themselves\cite{SnikKellerreview}.
The same set-up can be used in a thermally controlled environment to assess the errors due to thermal variations.
These errors may also include the effects of stress birefringence, which is unpredictable at this moment.
The effects of polarized fringes and other, yet unmodeled effects (e.g., dichroism) can be assessed as well by feeding the modulator with polarized light with a degree of polarization $< 10^{-3}$.

On sky, the polarimetric performance should be verified by observing polarized and unpolarized standard stars.
The polarized standard stars will be used to calibrate the continuum polarization, whereas stable magnetic stars that have already been observed with FORS\cite{FORSlegacy} can be used to assess issues with the measurement of line polarization (e.g., due to flexure\cite{Xshooterflexure} and variable grating illumination).
Instrumental polarization may also occur due to nonuniform coatings of M1 and/or M2.

\section{OPTICAL DESIGN}

\begin{figure}[t]
\centering
\includegraphics[width=0.45\textwidth]{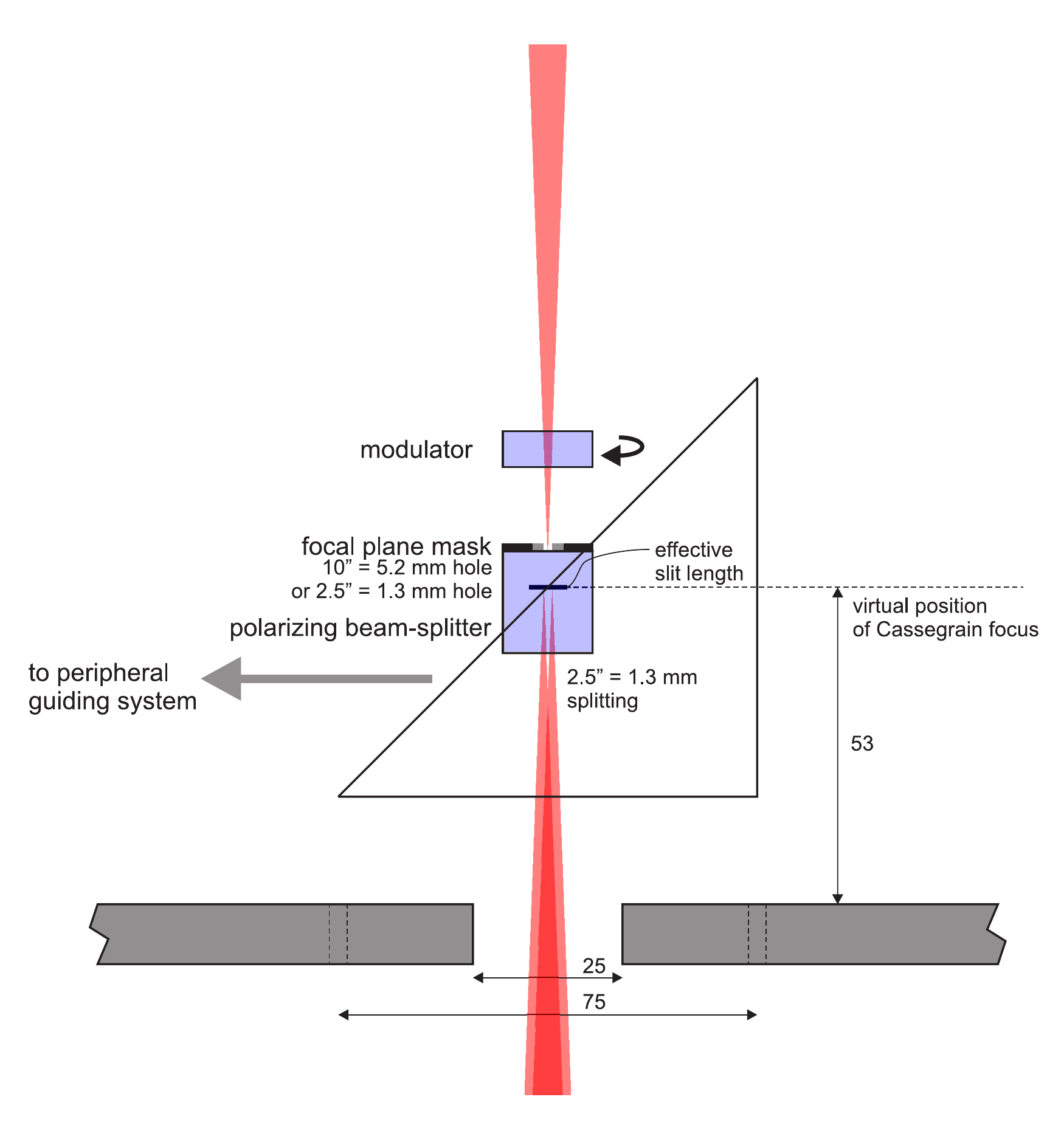}
\includegraphics[width=0.45\textwidth]{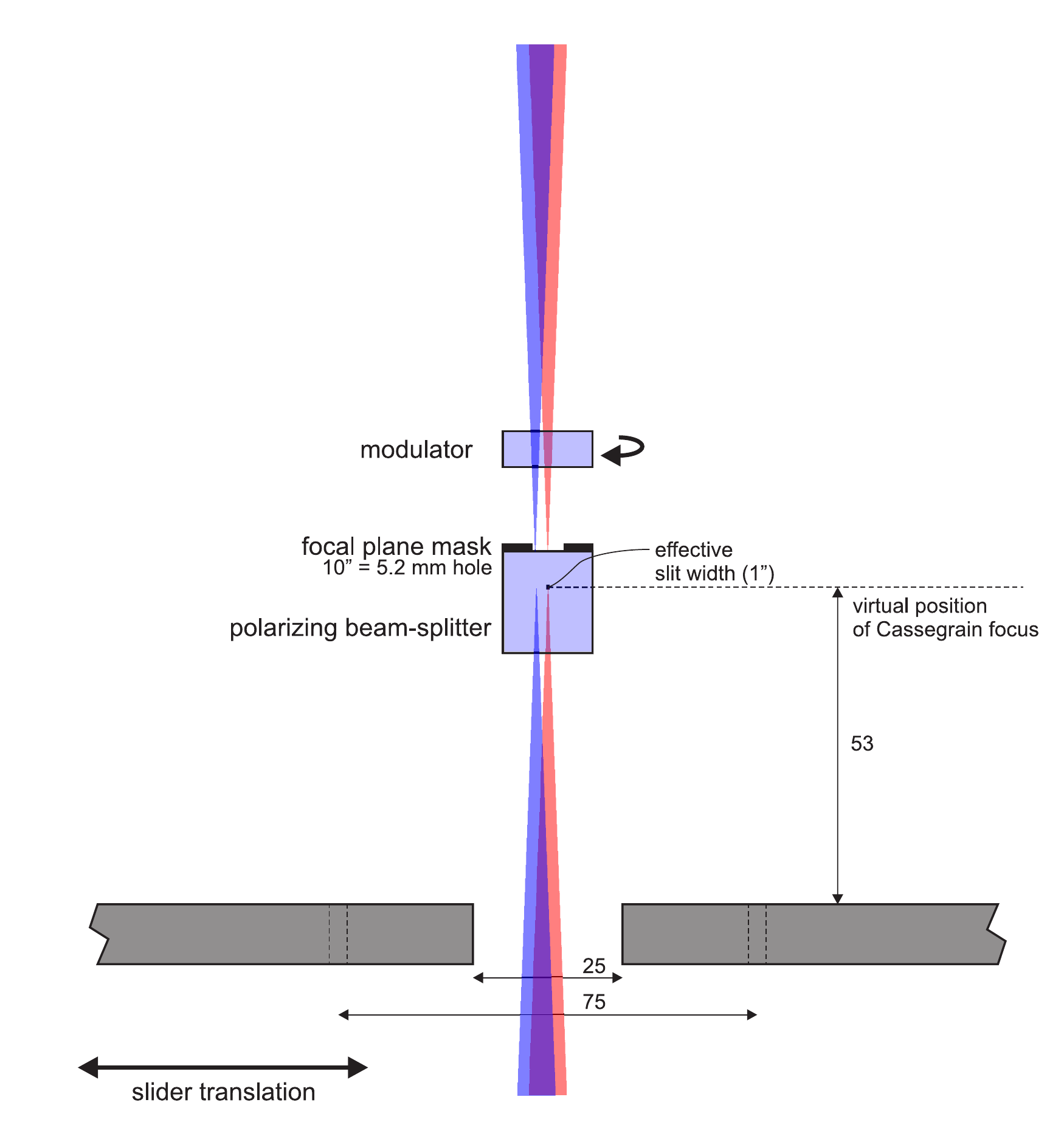}
\caption{Conceptual design for the X-shooter polarimeter unit. Two different side-views.}
\label{conceptualdesign}
\end{figure}

\begin{figure}[t]
\centering
\includegraphics[width=0.7\textwidth]{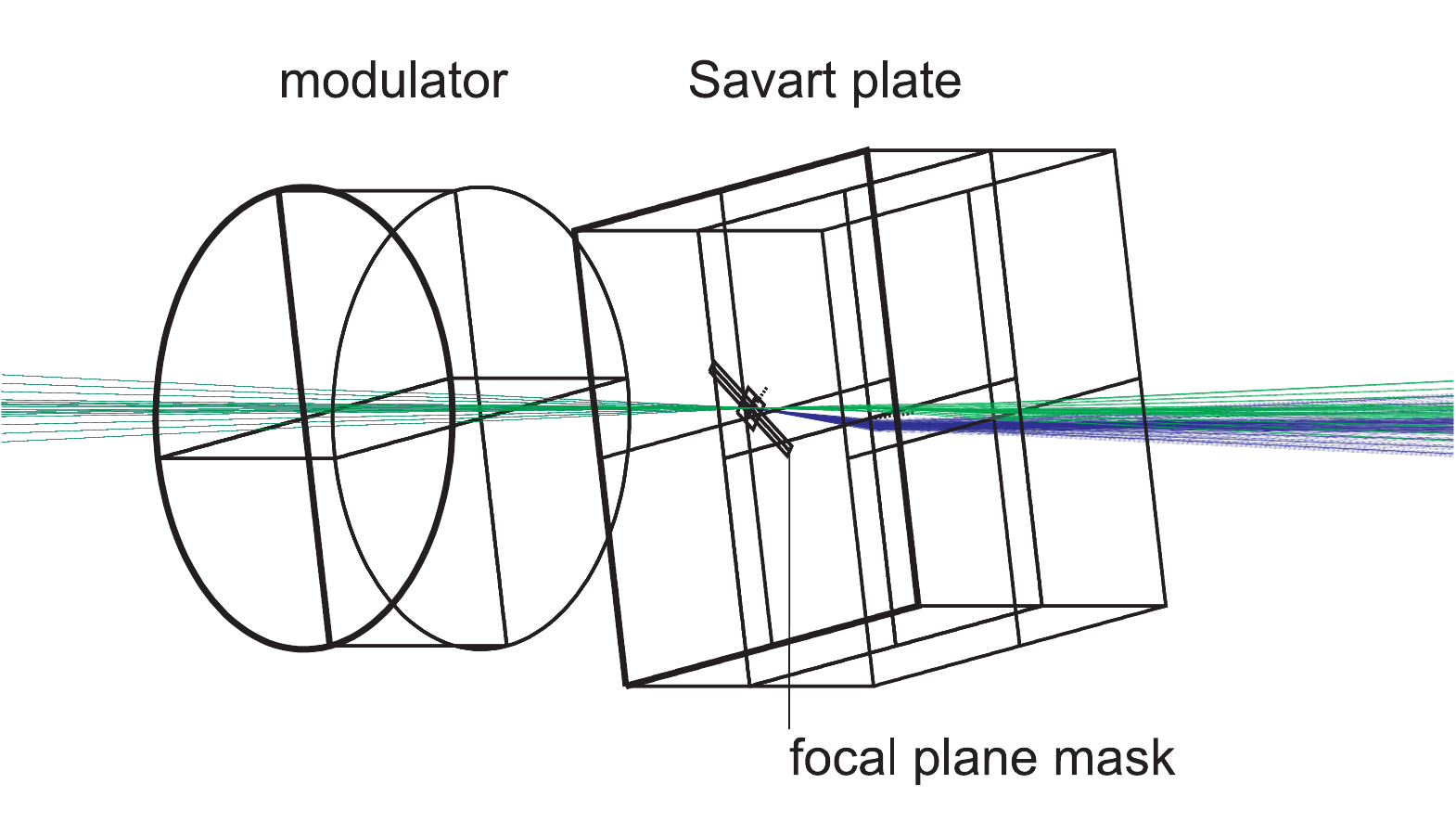}
\includegraphics[width=\textwidth]{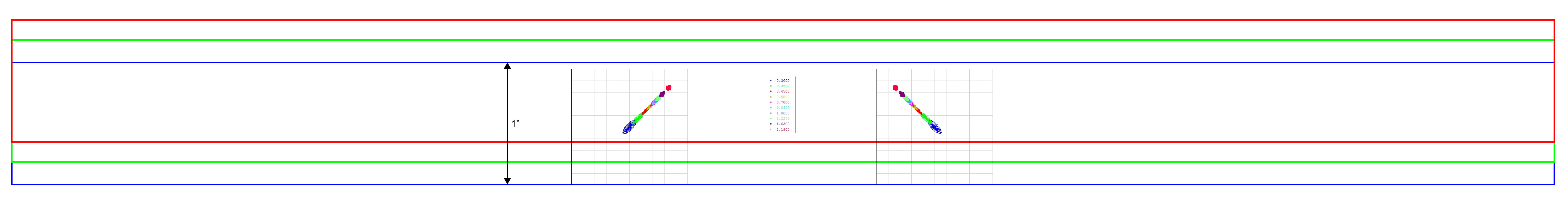}
\caption{Optical design for the X-shooter polarimeter unit. The bottom panel indicates the split beams as a function of wavelength for the three 1'' slits of the X-shooter spectrograph arms.}
\label{opticaldesign}
\end{figure}

The conceptual design of the X-shooter polarimeter unit is presented in Fig.~\ref{conceptualdesign}.
Taking the selected polychromatic modulator as a starting point, the optical design basically consisted of the design of the polarizing beam-splitter.
The separation of the beams needs to be along the slit direction (which is common to the three spectrographs of X-shooter).
The amount of splitting should be $\sim$2'', such that 1'' PSFs can be covered without significant overlap.

As the polarimeter is located in a focal plane, the beam-splitter of choice is a Savart plate\cite{SnikKellerreview}.
Such a device consists of one or two birefringent plates, cut at 45 degrees with respect to the entrance face. 
For the combination of two plates, the path lengths of the two beams are identical, and their aberrations are symmetrical.
The displacement between the two beams is then at 45$^\circ$ with respect to the split linear polarization directions.
The amount of splitting for a plate with a given thickness is determined by the birefringence of the material.
Given the limited amount of volume available for the polarimeter, the only viable birefringent material for the Savart plate is calcite.
Unfortunately, the e beam of calcite becomes opaque at 2.2 $\mu$m\cite{calciteIR}, and since the e and o beams are swapped in the two-plate Savart, the X-shooter polarimeter cannot be used for wavelengths upward of 2.2 $\mu$m.
Calcite of good quality is transparent down to 210 nm.

Savart plates are known to suffer from several aberrations when placed in converging beams: e.g., crystal astigmatism and spherochromatism.
However, the chromaticity of the splitting itself is by far the largest effect, see Fig.~\ref{opticaldesign}.
Fortunately, the chromatic splitting is smaller than the equivalent of 1'' (i.e.~the slit width), particularly when the alignments of all three slits are performed individually.

A $\sim$2'' focal plane mask is implemented in the Cassegrain focus above the Savart plate, to avoid overlap between the split point source under investigation an nearby structure.
A wider mask may be implemented that accommodates nodding in the IR spectrograph.
Because the Savart plate changes the optical properties of X-shooter as a whole, the Cassegrain focus needs to be physically moved upwards, such that this position matches the slit planes again.
Also taking into account the focus shift due to the modulator, a telescope needs to be defocused by moving M2 backwards by $\sim$0.16 mm.
As the Savart plate shifts the beams in addition to splitting them, offset pointing by $\sim$1.8'' w.r.t.~the nominal focus is required.

Because atmospheric dispersion correction is performed for all three spectrographs indicvidually, downstream of the Cassegrain optics, the polarimeter needs to accommodate an atmospherically dispersed beam.
And since the focal plane mask limits the FOV along the slit direction, this dispersion needs be aligned perpendicular to the slit direction, and the polarimeter's focal plane mask needs therefore to be opened up in that direction by $\sim$4'', see Fig.~\ref{conceptualdesign}.
This essentially fixes the orientation of X-shooter in polarimetric mode in one orientation of the Cassegrain rotator.
This has as a consequence that the $[Q,U]$ coordinate system on the sky will rotate during observations.
This cannot generally be corrected with offsets to the modulator angles, as the polychromatic modulator is an elliptical retarder.
Therefore, one modulation cycle always needs to be completed within the time after which the $[Q,U]$ rotation becomes a significant source of error.
Derotation can then be performed in data-reduction.
The largest rotation rates occur for targets that cross close to zenith. 
As atmospheric dispersion is minimal at those angles, X-shooter may then be corotated with the sky.

\section{CONCLUSIONS \& OUTLOOK}
The most challenging aspects of the design of the X-shooter polarimeter are the polarimetric implementation and the optical design.
The lights are now on green for these aspects, thanks to the implementation of a polychromatic modulator.
In the near future we will produce a prototype modulator to verify its performance in the lab.
The next step is to make progress on the mechanical design and implementation of the electronics/software for rotating the modulator.

We aim to bring together a large and versatile science team for the X-shooter polarimeter soon.
This way, we hope to make sure that all aspects that could limit the  performance regarding any of the scientific opportunities are dealt with before construction of the instrument.
Furthermore, we aim to construct a generic data-reduction pipeline during the early phases of the project, such that observing with the X-shooter polarimeter becomes as streamlined as possible.



\scriptsize{

\bibliography{Sniketal2012-Xshooterpol-SPIE_8446-76}

}

\end{document}